            \documentclass[preprint,12pt,authoryear]{elsarticle}

\usepackage{amssymb}
\usepackage{amsmath}
\usepackage{etoolbox}
\usepackage{amssymb}
\usepackage{graphicx}

\newcommand{\vect}[1]{\boldsymbol{#1}}

\newtheorem{definition}{\textbf{Definition}}

\newtheorem{proposition}{\textbf{Proposition}}

\newtheorem{corollary}{\textbf{Corollary}}
\newtheorem{lemma}{\textbf{Lemma}}

\usepackage{algorithm}

\usepackage{algorithmic}

\newtheorem{remark}{\textbf{Remark}}
\newtheorem{example}{\textbf{Example}}

\usepackage[none]{hyphenat}

\journal{Automatica}

\begin{document}

\begin{frontmatter}


\tnotetext[label1]{This research work was supported by the project COGNIMUSE which is implemented under the ARISTEIA Action of the Operational Program Education and Lifelong Learning and is co-funded by the European Social Fund
	and Greek National Resources. The work of G.P. Papavassilopoulos was also supported by the program ARISTEIA, project name HEPHAISTOS}
\title{Stochastic Stability in Max-Product and Max-Plus Systems with Markovian Jumps\tnoteref{label1}}


\author[label_addr1]{Ioannis Kordonis}
\author[label_addr]{Petros Maragos}
\author[label_addr]{George P. ~Papavassilopoulos}
\address[label_addr1]{CentraleSupl\'ec, 
Avenue de la Boulaie, 35576 Cesson-S\'evign\'e, France. }
\address[label_addr]{National Technical University of Athens, School of Electrical and Computer Engineering, 9 Iroon Polytechniou str., Athens, Postal Code 157 80, Greece}

\address{}

\begin{abstract}
 We study Max-Product and Max-Plus Systems with Markovian Jumps and focus on stochastic stability problems. At first, a Lyapunov function is derived for the asymptotically stable deterministic Max-Product Systems.  This Lyapunov function is then adjusted to derive sufficient conditions for the stochastic stability of Max-Product systems with Markovian Jumps. Many step Lyapunov functions are then used to derive necessary and sufficient conditions for stochastic stability. The results for the Max-Product systems are then applied to Max-Plus systems with Markovian Jumps, using an isomorphism and almost sure bounds for the asymptotic behavior of the state are obtained. A numerical example illustrating the application of the stability results on a production system is also given. 
\sloppy
\end{abstract}


\end{frontmatter}


 \section{Introduction}
 \sloppy
 Max-Plus systems are dynamical systems which satisfy the superposition principle in the Max-Plus algebra. The use of Max-Plus systems was proposed in various applications involving timing, such as communication and traffic management, queueing systems, production planning, multi-generation energy systems, et.c. (eg. \cite{cuninghame1979minimax}, \cite{baccelli1992synchronization}, \cite{heidergott2014max}, \cite{goverde2007railway}, \cite{baccelli2000tcp}). Recently, the use of the closely related class of Max-Product systems (systems which satisfy the superposition principle in the Max-Product algebra) was proposed  as a tool for the modelling  of cognitive processes, such as detecting audio and visual salient events in multimodal video streams (\cite{maragos201}). Max-Plus and Max-Product algebras have also computational uses  involving Optimal Control problems (\cite{mceneaney2006max}) and estimation problems in probabilistic models such as the max-sum algorithm in Probabilistic Graphical models and the Viterbi algorithm in Hidden Markov Models (eg. \cite{bishop2006pattern}).
 
 In this work, we study stochastic Max-Plus and Max-Product systems, where the system matrices depend on a finite state Markov chain. For the Max-Plus systems we focus on the asymptotic growth rate, whereas for the Max-Product systems on stochastic stability.   A motivation to study Max-Plus systems with Markovian jumps is to model production systems, where the processing or holding times are  random variables (not necessarily independent) or there are random failures and repairs, modeled as a Markov chain. The results on max-product stochastic systems will be used as an intermediate step. An independent  motivation to study Max-Product systems is the modeling of cognitive processes interrupted by random events. 
 Similar problems with Markovian delays for linear systems were
 studied  in  \cite{beidas1993convergence}, for random failures in \cite{papavassilopoulos1994distributed}  and  for  nonlinear  time 
 varying  systems  in \cite{beidas1995distributed},  in  the  context  of  distributed parallel optimization and  routing applications. In the current work, we try to exploit the special (Max-Product or Max-Plus) structure of the system. 
 
 At first, deterministic Max-Product systems are considered and their asymptotic stability is characterized using Lyapunov functions. The  Lyapunov function derived can be also used to study systems which are not linear in the Max-Product algebra. We then study Max-Product systems with Markovian Jumps and derive sufficient conditions for their stochastic stability. Further, necessary and sufficient conditions for the stochastic stability of Max-Product systems with Markovian Jumps are derived using many step Lyapunov functions.
 The results for the stochastic stability of Max-Product systems are then used to derive bounds for the evolution of the state of Max-Plus systems with Markovian Jumps.
 
 The results of this work relate to the literature for the approximation of the Lyapunov exponent of Max-Plus stochastic systems. The existence of the Lyapunov exponent was proved in \cite{cohen1988subadditivity}. Limit theorems for the scaled asymptotic evolution of stochastic Max-Plus systems were proved in \cite{merlet2007central}, \cite{merlet2008cycle}. Most of the works on the approximation of the Lyapunov exponent focus on the independent random matrix case. In \cite{baccelli2000analytic}, \cite{gaubert2000series} series expansions are used in order to approximate the Lyapunov exponent and \cite{goverde2008fast}, \cite{goverde2011coupling} use approximate stochastic simulation techniques to estimate the Lyapunov exponent. In \cite{blondel2000approximating} it is shown that the approximation of the Lyapunov exponent is an NP-hard problem.  Bounds for the tail distributions of Max-Plus stochastic systems are proposed in \cite{chang1996exponentiality}. In \cite{liu1995bounds}, a model of Max-Plus system with Markovian input is considered and bounds for the tail distributions are derived. A model where the Markov chain (branching process) evolves according to a Max-Plus stochastic system is analyzed in \cite{altman2012branching}. Bounds on the length of the transient phase of Max-Plus systems are proved in \cite{nowak2014overview}.
 
 Another related class of systems is Switching Max-Plus systems with deterministic or stochastic switching introduced in \cite{van2006modelling} and studied further in \cite{van2012modeling}. The basic difference with the current work is that the current work focuses on stochastic stability properties whereas \cite{van2006modelling}, \cite{van2012modeling} study stability under arbitrary switching. Several approximation methods in stochastic Max-plus systems control and identification were studied in \cite{safaei2012approximation}.

The techniques used in this work closely parallel the techniques used for the stability analysis of Markovian Jump Linear Systems (MJLS). The study of the stochastic stability of MJLS dates back at least to the 1960’s (\cite{Bharucha_PhD}) and today is a well-established field   (eg. \cite{costa2006discrete}, \cite{Fang_Loparo}, \cite{beidas1993convergence}, \cite{papavassilopoulos1994distributed}, \cite{kordonis2014stability}).

 \subsection{Background}
 
 The Max-Plus and Max-Product algebras are used. In the Max-Plus algebra the usual summation is substituted by maximum and the usual multiplication is substituted by summation. In the Max-Product algebra the usual summation is substituted by maximum but the multiplication remains unchanged. 
 
 The Max-Plus algebra is defined on the set of extended reals $\mathbb{\bar R}=\mathbb{R}\cup\{-\infty,+\infty\}$ with the binary operations ``$\vee$'' and ``$\boxplus$".  The operation ``$\vee$" stands for the maximum i.e., for $x,y\in\mathbb{\bar R}$, it holds $x\vee y = \text{max} \{x,y\}$. The operation ``$\boxplus$" corresponds to the usual addition i.e.,  for  $x,y\in\mathbb{\bar R}$ it holds $x\boxplus y =x+y$, where  the convention $-\infty \boxplus\infty=-\infty$ is used. For a set  $(x_i)_{i\in I}$ of extended reals ``$\bigvee $" stands for the supremum i.e. $\bigvee_{i\in I} x_i=\sup_{i\in I} \{x_i\}$. For a pair of matrices $\boldsymbol{A} =[A_{ij}]$ and $\boldsymbol{B}=[B_{ij}]$, the operation ``$\vee$" is their element-wise maximum, i.e.:
 \begin{equation}
 (\boldsymbol A\vee\boldsymbol B)_{ij} = A_{ij}\vee B_{ij},\nonumber
 \end{equation}
 and similarly is the element-wise supremum for an arbitrary set of matrices. 
 
 For a pair of matrices $\boldsymbol A=[A_{ij}]\in \mathbb{\bar R}^{n\times m}$ and $\boldsymbol B=[B_{ij}]\in  \mathbb{\bar R}^{m\times l}$ their Max-Plus product $\boldsymbol A\boxplus \boldsymbol B $ is an $n\times l$ matrix and its $i,j$-th element is given by:
 \begin{equation}
 (\boldsymbol A\boxplus \boldsymbol B)_{ij}= \bigvee_{p=1}^m \left(A_{ip}+B_{pj} \right), \label{nota1}
 \end{equation}
 where ``$\bigvee$" denotes the maximum of the $m$ elements.
 
 The Max-Product algebra is defined on $\mathbb{\bar R}_+=[0,\infty]$, with the binary operations ``$\vee$" and ``$\boxtimes$". The ``$\boxtimes$" operation is the usual scalar multiplication with the convention $0\boxtimes\infty=0$. The ``$\vee$" operation is defined exactly as in the Max-Plus algebra. The matrix multiplication in the Max-Product algebra is defined by:
 \begin{equation}
 [\boldsymbol A\boxtimes \boldsymbol  B ]_{ij}= \bigvee_{p=1}^m \left(A_{ip}B_{pj} \right),\nonumber
 \end{equation}
 
 The power of a square matrix is defined by $\boldsymbol A^k=\boldsymbol A^{k-1}\boxtimes \boldsymbol A$ and $\boldsymbol A^0=\boldsymbol I$.
 For a given square matrix $\boldsymbol A$ a new matrix $\boldsymbol A^+$ is defined as $\boldsymbol A^+=\bigvee_{k=0}^\infty \boldsymbol A^k $.
 The subset $\mathbb{R}_+=[0,\infty)$ of $\mathbb{\bar R}_+$ will be also used.
 
 Max-Product multiplication distributes over ``$\bigvee$", i.e.: 
 \begin{align}
 \bigvee_{i\in I} \boldsymbol A\boxtimes\boldsymbol  B_i =\boldsymbol A\boxtimes \left(\bigvee_{i\in i}  \boldsymbol B_i \label{nota2}\right).
 \end{align} 
 The same property holds also for the Max-Plus multiplication.
 
 In both algebras, the ``$\vee$" operation has lower priority than ``$+$" or ``$\boxplus$'' in the Max-Plus algebra and ``$\cdot$" or ``$ \boxtimes$'' in the Max-Product algebra respectively.
 Let us note that there is an isomorphism $\exp(\cdot)$ between the Max-Plus algebra $(\mathbb{\bar R},\vee,\boxplus)$ and the  Max-Product algebra $(\mathbb{\bar R}_+,\vee,\boxtimes)$.
 
 A unifying algebraic framework to study Max-Plus and Max- Product systems  (and also other systems) is the theory of Weighted Lattices (\cite{maragos2013}, \cite{Maragos_submited}). 

\begin{table}
\label{Notation}
\centering 
\begin{tabular}{c | c }
Operation & Meaning \\
\hline 
$\vee$  & The maximum. Applies for scalars, vectors and matrices\\
$\boxplus$  & Max-plus multiplication. Defined in \eqref{nota1}\\
$\boxtimes$  & Max-plus multiplication. Defined in \eqref{nota2}\\
\end{tabular}
\caption{The algebraic operations used. } 
\label{table:notationf} 
\end{table}

 \subsection{Notation}
 
 For a pair of vectors $\boldsymbol x=(x_1,\dots,x_n)^T$ and $\boldsymbol y=(y_1,\dots,y_n)^T$, the inequality notation $\boldsymbol x\leq \boldsymbol y$ is used meaning that $x_i\leq y_i$, for all  $i$. Similarly, the inequality notation $\boldsymbol x<\boldsymbol y$ stands for $x_i<y_i$, for  all  $i$. The infinity norm will be used i.e. $\|\boldsymbol x\|=\max_i |x_i|$. We denote by $\vect 1$  the column vector of dimension $n$ consisting of ones.
 The underlying probability space is denoted by $(\Omega,\mathcal{F},P)$.
 
 A function $\alpha:\mathbb{R}_+\rightarrow\mathbb{R}_+$ will be called class $\mathcal{K}$ function if $\alpha$ is increasing and $\alpha(0)=0$. A function  $\beta:\mathbb{R}_+\times\mathbb{R}_+\rightarrow\mathbb{R}_+$ will be called class $\mathcal{KL}$ function if, for each fixed $t$, the function $\beta(\cdot,t)$ is a class $\mathcal{K}$ function and for any fixed $s$, the function  $\beta(s,\cdot)$ is decreasing and $\beta(s,t)\rightarrow 0$ as $t\rightarrow\infty$.
 
 \subsection{Problem Formulation}
 
 The first class of systems considered is Max-Product systems with Markovian jumps. The uncertainty of the system is described by a Markov chain $y_k$ having a finite state space $\{1,\dots,M\}$ and transition probabilities $c_{ij}$. That is, the evolution of $y_k$ is described by $c_{ij}=P(y_{k+1}=j|y_k=i)$. A Max-Product system with Markovian jumps is described by:
 \begin{align}
 \boldsymbol x_{k+1}&=\boldsymbol A(y_k)\boxtimes \boldsymbol x_k\label{Max_Prod_MJ},\\
 \boldsymbol x_0&\in \mathbb{ R}_+^n \nonumber.
 \end{align}
 That is, at each time step the system matrix $A$  takes one of the $M$  different values $A(1),\dots,A(M)$ according to the value of the Markov chain.

 At first, the class of deterministic Max-Product systems will be considered. In these systems the matrix $\boldsymbol A(\cdot)$ does not depend on the Markov chain and takes a single value $\boldsymbol A$.
 
 The other class of systems considered is Max-Plus systems with Markovian jumps in the form:
 \begin{align}
 \boldsymbol x_{k+1}&=\boldsymbol A(y_k)\boxplus \boldsymbol x_k\label{Max_Plus_MJ},\\
 \boldsymbol x_0&\in \mathbb{  R}^n \nonumber.
 \end{align}

 In the following definitions, some notions of stability and stochastic stability are recalled from the literature (eg. \cite{khalil2002nonlinear}, \cite{Maragos_submited} and \cite{kozin1969}).
 
 \begin{definition}
 	Consider the system:
 	\begin{align}
 	\boldsymbol x_{k+1}&=\left(\boldsymbol A\boxtimes \boldsymbol x_k \right)\vee \left(\boldsymbol B\boxtimes \boldsymbol u_k\right), \label{Det_sys_with_inp} \\
 	\boldsymbol z_k &=\left(\boldsymbol  C\boxtimes \boldsymbol x_k \right)\vee \left(\boldsymbol D\boxtimes \boldsymbol u_k\right), \label{Det_sys_with_inp_outp} 
 	\end{align} 
 	where $\boldsymbol x_k$,  $\boldsymbol u_k$,  $\boldsymbol z_k$, denote the system state, input and output and $A, B,C,D$ are matrices of appropriate dimensions. 
 	
 	\begin{itemize}
 		\item[(i)] The free system, i.e. \eqref{Det_sys_with_inp} with $\boldsymbol u_k=0$, is exponentially stable, if there exist constants $a>1$ and $L>0$ such that
 		$\|x_k\|\leq L\|x_0\|/a^k$,  for any initial conditions and any $k$.
 		\item[(ii)] The system \eqref{Det_sys_with_inp} is Input to State Stable (ISS) if  there exist a class $\mathcal{KL}$ function $\beta$ and a class $\mathcal{K}$ function $\alpha$ such that:
 		\begin{align*}
 		\|\boldsymbol x_k\|\leq \beta (\|\boldsymbol x_0\|,k)+\alpha\left(  \bigvee_{i=0}^k \|\boldsymbol u_k\|\right),
 		\end{align*}
 		for any initial condition, any $k$ and any input sequence $\boldsymbol u_k$.
 		\item[(iii)] The system \eqref{Det_sys_with_inp}, \eqref{Det_sys_with_inp_outp} is Bounded Input Bounded Output (BIBO) stable (\cite{Maragos_submited}) if $\bigvee_{k=0}^\infty \|\boldsymbol  u_k\|<\infty$ implies $\bigvee_{k=0}^\infty \|\boldsymbol z_k\|<\infty$, for any initial conditions.  
 	\end{itemize}
 \end{definition}
 
 \begin{definition}
 	The system given by \eqref{Max_Prod_MJ} is:
 	\begin{itemize}
 		\item[(i)] Almost surely stable if for any initial conditions, $\boldsymbol x_k\rightarrow 0$ almost surely.
 		\item[(ii)] Mean norm stable if  $E[\|\boldsymbol x_k\|]\rightarrow 0$ as $k\rightarrow \infty$.
 		\item[(iii)] \label{Part3Def1}
 		Mean norm exponentially stable if there exist constants $a>1$ and $L>0$ such that    $E[\|\boldsymbol x_k\|]\leq L\|\boldsymbol x_0\|/a^k$.
 	\end{itemize}
 \end{definition}
 
 Conditions for the stochastic stability of systems in the form \eqref{Max_Prod_MJ} will be derived. For Max-Plus systems bounds on the growth of $\boldsymbol x_k$ will be derived.


 \section{Deterministic Max-Product Systems}
 \label{Deterministic_Max_Product}
 
 In this section the asymptotic stability of deterministic Max-Product systems in the form:
 \begin{align}
 \label{Det_sys}
 \boldsymbol x_{k+1}&=\boldsymbol A\boxtimes \boldsymbol x_k, \\ \nonumber
 \boldsymbol x_0&\in \mathbb{R}_+^n,
 \end{align}
 is studied. 
 
The following Lemma presents a condition equivalent to the the exponential stability   of \eqref{Det_sys} (for a definition of exponential stability see for example \cite{khalil2002nonlinear}).
 
 \begin{lemma}
 	\label{Properties_Lemma}
 	It holds:
 	\begin{itemize}
 		\item[(i)] The function $f(\boldsymbol x)=\boldsymbol A\boxtimes \boldsymbol x$ is homogeneous of order $1$, i.e. it holds $f(\rho \boldsymbol x)=\rho f(\boldsymbol x)$ for any $\rho \in\mathbb{R}_+$.
 		\item[(ii)] The system \eqref{Det_sys} is exponentially stable iff for some $a>1$, the system $\boldsymbol x_{k+1}=a\boldsymbol A\boxtimes \boldsymbol x_k$ is stable.
 	\end{itemize}
 \end{lemma}
 \textit{Proof:} The proof is immediate. \hfill $\square$
 
 A Lyapunov function will be constructed for the stable systems in the form \eqref{Det_sys}. Consider the function:
 \begin{equation}
 V(\boldsymbol x)=\bigvee_{k=0}^\infty \boldsymbol \lambda^T \boxtimes \boldsymbol A^k \boxtimes \boldsymbol x,
 \label{Lyap_Det}
 \end{equation}
 where $\boldsymbol \lambda$   is a vector with positive entries. Equivalently, $V$ can be written as 
 \begin{equation}
 V(\boldsymbol x)=\bigvee_{k=0}^\infty \boldsymbol \lambda^T \boxtimes \boldsymbol x_k,\nonumber
 \end{equation}
 where $\boldsymbol x_k$ is the state vector of \eqref{Det_sys} with initial condition $\boldsymbol x_0=\boldsymbol x$.
 It is not difficult to see that if \eqref{Det_sys} is stable, then $V(\boldsymbol x)$ is finite for any $\boldsymbol x$  and $V(\boldsymbol 0)=0$. Furthermore, the sequence $V(\boldsymbol x_k)$ is non-increasing:
 \begin{equation}
 \bigvee_{k=k_0+1}^\infty \boldsymbol \lambda^T \boxtimes \boldsymbol x_k \leq\bigvee_{k=k_0}^\infty \boldsymbol \lambda^T \boxtimes \boldsymbol x_k \nonumber.
 \end{equation}
 Thus, $V$ is a Lyapunov function. 	
 
 The form of $V$ can be computed using the following calculations:
 \begin{align}
 V(\boldsymbol x)&=\bigvee_{k=0}^\infty \boldsymbol \lambda^T \boxtimes \boldsymbol A^k\boxtimes \boldsymbol x=\boldsymbol \lambda^T \boxtimes\left[\bigvee_{k=0}^\infty  \boldsymbol A^k\right]\boxtimes \boldsymbol x\nonumber \\
 &= (\boldsymbol \lambda^T\boxtimes  \boldsymbol A^+)\boxtimes \boldsymbol x \label{LYap_form_min1}.
 \end{align}
 Thus, $V$ has the form:
 \begin{equation}
 V(\boldsymbol x)=\boldsymbol p^T\boxtimes \boldsymbol x, \label{LYap_form}
 \end{equation}
 where $\boldsymbol p$ is an $n$ vector with positive entries.
 
 \begin{proposition}
 	The following are equivalent:
 	\begin{itemize}
 		\item[(i)] The system \eqref{Det_sys} is exponentially stable.
 		\item[(ii)] There exists a vector $\boldsymbol p$, with positive entries, such that $\boldsymbol A^T\boxtimes \boldsymbol p<\boldsymbol p$
 	\end{itemize}
 \end{proposition}
 
 \textit{Proof:} 
 In order to show the direct part, we use Lemma \ref{Properties_Lemma}, to obtain a constant $a>1$ such that $\boldsymbol x_{k+1}=a\boldsymbol A\boxtimes \boldsymbol x_k$ is stable. Using a Lyapunov function in the form \eqref{Lyap_Det} for that system, we obtain a positive vector $\boldsymbol p$ such that $V(\boldsymbol x)=\boldsymbol p^T\boxtimes \boldsymbol x$. Then it holds:
 \begin{equation}
 a\boldsymbol p^T\boxtimes \boldsymbol A\boxtimes \boldsymbol x\leq \boldsymbol p^T\boxtimes \boldsymbol x \nonumber,
 \end{equation}
 for any $\boldsymbol x\in \mathbb{R}_+^n$. Thus, $\boldsymbol p^T\boxtimes \boldsymbol A <\boldsymbol p^T$ or equivalently  $\boldsymbol A^T\boxtimes \boldsymbol p<\boldsymbol p$.
 
 The fact that (ii) implies (i) is shown with usual Lyapunov analysis. \hfill $\square$
 
 \begin{remark}
 	The condition $a\boldsymbol p^T\boxtimes \boldsymbol A\leq \boldsymbol p^T$ can be checked using Linear Programming.  
 \end{remark}
 
 \begin{remark}
 	A Lyapunov function in the form \eqref{Lyap_Det} is the direct analogue of a Lyapunov function for a usual linear system $\boldsymbol x_{k+1}=\boldsymbol A\boldsymbol x_k$, $\boldsymbol x_0=x$ in the form $V_L(\boldsymbol x)=\sum_{k=0}^\infty \boldsymbol x_k^T\boldsymbol Q\boldsymbol x_k $. Particularly, in the place of the summation, we have the supremum and in the place of the $\boldsymbol Q$-norm $\|\boldsymbol x\|_{\boldsymbol Q}^2 = \boldsymbol x^T\boldsymbol Q\boldsymbol x$ we have the $\boldsymbol \lambda$-norm $\|\boldsymbol x\|_{\boldsymbol \lambda} = \text{max} \{\lambda_1x_1,\dots,\lambda_nx_n \}$. 
 \end{remark}
 
 \begin{remark}
 	The asymptotic behaviour of Max-Plus deterministic systems, depends on the  Max-Plus eigenvalue of the system matrix which under connectivity assumptions turns out to be unique (eg. \cite{baccelli1992synchronization}). This eigenvalue can be computed in terms of the critical paths i.e. the paths with maximal average weight.  
 	This analysis can be transferred to Max-Product systems using the $\exp(\cdot)$ isomorphism of the Max-Plus and Max-Product algebras. The Lyapunov approach adopted here could, however, be extended to stochastic systems and systems which are not linear in the Max-Product algebra. 
 \end{remark}

 The following corollary studies the Input to State Stability (ISS)  and the Bounded Input Bounded Output (BIBO) stability. 
 \begin{corollary}
 	Assume that the system given by \eqref{Det_sys} is exponentially stable. Then:
 	\begin{itemize}
 		\item[(i)] The system given by \eqref{Det_sys_with_inp} is input to state stable.
 		\item[(ii)] The system given by \eqref{Det_sys_with_inp}, \eqref{Det_sys_with_inp_outp}  is BIBO stable.
 	\end{itemize}
 \end{corollary}
 \textit{Proof}: (i) Consider a Lyapunov function $V$ in the form \eqref{LYap_form}. Then $V$ satisfies Lemma 3.5 of  \cite{jiang2001input}. Thus, the system is ISS. 
 
 (ii) Follows immediately from (i).\hfill $\square$

 The following example illustrates that the same Lyapunov functions can be used to analyze systems  which are nonlinear in the Max-Product algebra.
 \begin{example}
 	Consider the system:
 	\begin{equation}
 	\boldsymbol x_{k+1}=\left[\begin{matrix}
 	2/3& 2\\
 	1/3& 3/4
 	\end{matrix} \right] \boxtimes \boldsymbol x_k. \label{Example1_DYn}
 	\end{equation}
 	We consider the Lyapunov function candidate $V(\boldsymbol x)=[2 ~5]\boxtimes \boldsymbol x$. It holds:
 	\begin{equation}
 	[2 ~5]\left[\begin{matrix}
 	2/3& 2\\
 	1/3& 3/4
 	\end{matrix} \right] =[5/3 ~4]<[2~5] \nonumber.
 	\end{equation}
 	Thus, $V$ is a Lyapunov function and the system \eqref{Example1_DYn} is exponentially stable. 
 	
 	Let us then consider the system:
 	\begin{equation}
 	~~~~~~~\boldsymbol x_{k+1}=\left[\begin{matrix}
 	2/3& 2\\
 	1/3& 3/4
 	\end{matrix} \right] \boxtimes \boldsymbol x_k\vee \left[\begin{matrix}
 	2\\
 	3
 	\end{matrix} \right]\boxtimes (\boldsymbol x_k^T\boxtimes\boldsymbol  x_k ),\label{non_lin1}
 	\end{equation}
 	which is not in the form of \eqref{Det_sys}. The same Lyapunov function $V(\boldsymbol x)$ can be used to show the local asymptotic stability of \eqref{non_lin1}. 
 	
 	Furthermore, the same Lyapunov function $V(\boldsymbol x)$ can be used to show the ISS of the system:
 	\begin{equation}
 	~~~~~~~~~~~~\boldsymbol x_{k+1}=\left[\begin{matrix}
 	2/3& 2\\
 	1/3& 3/4
 	\end{matrix} \right] \boxtimes \boldsymbol x_k\vee \left[\begin{matrix}
 	5\\
 	8
 	\end{matrix} \right]\boxtimes u_k.\nonumber 
 	\end{equation}
 \end{example}

 \section{Max-Product Systems with Markovian Jumps}
 \label{Section_MAX_PROD_MJ}
 We then turn to Max-Product systems with Markovian Jumps in the form \eqref{Max_Prod_MJ}. Lyapunov functions in the form:
 \begin{equation}
 V(\boldsymbol x,y)=\boldsymbol p(y)^T\boxtimes \boldsymbol x,
 \label{Lyap_st1}
 \end{equation}
 generalizing \eqref{LYap_form} are considered. 
 
 \begin{proposition}
 	\label{Prop_lyap_Cond}
 	Assume that there exist a constant $a>1$ and  vectors with positive entries $\boldsymbol p(1),\dots,\boldsymbol p(M)$ such that:
 	\begin{equation}
 	a\sum_{j=1}^M c_{ij} \boldsymbol p(j)^T\boxtimes \boldsymbol A(i) \boxtimes \boldsymbol v \leq \boldsymbol p(i)^T\boxtimes \boldsymbol v,
 	\label{St_Lyap_cond1}
 	\end{equation} 
 	for any vector $\boldsymbol v$ with positive entries. Then, \eqref{Max_Prod_MJ} is mean norm exponentially stable and almost surely stable.
 \end{proposition}
 
 \textit{Proof:} Consider the function \eqref{Lyap_st1}. It holds:
 \begin{equation}
 E[V(\boldsymbol x_{k+1},y_{k+1})|\boldsymbol x_k=\boldsymbol x,y_k=i] =  \sum_{j=1}^M c_{ij} \boldsymbol p(j)^T\boxtimes \boldsymbol A(i)\boxtimes \boldsymbol x. \nonumber
 \end{equation}
 Condition \eqref{St_Lyap_cond1} implies that $V$  is a positive super-martingale. Furthermore, $V=0$ implies $\boldsymbol x=0$. Thus, the system is almost surely stable. 
 
 Condition \eqref{St_Lyap_cond1} further implies that:
 \begin{equation}
 E[V(\boldsymbol x_{k+1},y_{k+1})|V(\boldsymbol x_k,y_k)]\leq V(\boldsymbol x_k,y_k)/a. \nonumber
 \end{equation}  
 Thus, using this inequality repeatedly and taking expectations in both sides we have:
 \begin{equation}
 E[V(\boldsymbol x_{k},y_{k})]\leq V(\boldsymbol x_0,y_0)/a^k. \nonumber
 \end{equation}  
 Denoting by $p_M$ and $p_m$   the maximum and the minimum entry of  $\boldsymbol p(1),\dots,\boldsymbol p(M)$, we obtain:
 \begin{equation}
 E[p_m \|\boldsymbol x_k\|]\leq p_M\|\boldsymbol x_0\|/a^k. \nonumber
 \end{equation}  
 Thus, using $L=p_M/p_m$, the inequality in Definition \ref{Part3Def1} part (iii) holds and the system \eqref{Max_Prod_MJ} is mean norm exponentially stable. \hfill $\square$

 Condition \eqref{St_Lyap_cond1} should hold for any $\boldsymbol v\in\mathbb{R}_+^n$ and thus, it could be difficult to check it in general. The following lemma may be used to simplify condition \eqref{St_Lyap_cond1}. 
 The lemma will be used also in Section \ref{Many_Step_Lyapunov_Functions} which considers many step Lyapunov functions. Hence, the lemma will be stated using a possibly different timing (with $t$ in the place of $k$), a possibly different set of system matrices, depending on an additional random variable $w_t$ and a state vector $\bar x$ in the place of $x$.

 \begin{lemma}
 	\label{Lemma_1point}
 	Consider a system in the form:
 	\begin{equation}
 	\bar {\boldsymbol x}_{t+1} =  \bar{\boldsymbol A}(y_t,w_t)\boxtimes \bar {\boldsymbol x}_t,
 	\label{x_bar_equation}
 	\end{equation}
 	where $y_t$ takes values in $\{1,\dots,M\}$ and $w_t$ take values in $\{1,\dots,\bar M\}$. Assume also that $(y_{t},w_t)$ is a Markov chain and that $w_{t}$ is independent of $(w_{t-1}, y_{t-1})$ given $y_t$. Denote by $\tilde c(i,j,i')$ the conditional probability $P(y_{t+1}=i',w_t=j|y_t=i)$. Consider also the function: 
 	\begin{equation}
 	V(\bar {\boldsymbol x},y) = \boldsymbol p(y)^T\boxtimes \bar {\boldsymbol x},
 	\end{equation}
 	with $\boldsymbol p(1),\dots,\boldsymbol p(M)$ vectors with positive entries. For some $\delta$ with $0 < \delta < 1$, the following are equivalent:
 	\begin{itemize}
 		\item[(i)] It holds \begin{equation}
 		 E[V(\bar {\boldsymbol x}_{t+1},y_{t+1})|\bar {\boldsymbol x}_t,y_t]\leq \delta V(\bar {\boldsymbol x}_t,y_t),
\label{eqUl1}
 		 \end{equation} for all $\bar {\boldsymbol x}_t,y_t$.
 		\item[(ii)] It holds:
 		\begin{align}
 		\sum_{j=1}^{\bar{M}}  \sum_{i'=1}^M \tilde{c}(i,j,i') \vect 1^T \boxtimes \tilde{\boldsymbol A}(i',i,j) \boxtimes \vect 1 \leq \delta \label{Modif_cond},
 		\end{align}
 		for $i=1,\dots,M$, where 
 		\begin{align}
 		\tilde{\boldsymbol A}(y_{t+1},y_t,w_t)=\text{diag}(\boldsymbol p(y_{t+1})) \bar {\boldsymbol A}(y_t,w_t) \text{diag}(\boldsymbol p(y_{t})^{-1}).
 		\label{Alpha_tilda_eq}
 		\end{align}
 	\end{itemize}
 \end{lemma}
 
 \textit{Proof:} Consider the vector: 
 \begin{align*}
 \boldsymbol z_t=\text{diag}(\boldsymbol p(y_t))\boxtimes \bar {\boldsymbol x}_t.
 \end{align*} 
 Then, it holds:
 \begin{align}
 V(\bar{\boldsymbol x}_t,y_t) = \vect 1 ^T \boxtimes \boldsymbol z_t = \|\boldsymbol z_t\|.
 \label{gamisemas}
 \end{align} 
 Furthermore, $\boldsymbol z_t$ evolves according to:
 \begin{align*}
 \boldsymbol z_{t+1}=\tilde{\boldsymbol A}(y_{t+1},y_t,w_t) \boxtimes \boldsymbol z_t.
 \end{align*}

 Let us first show that (i) can be expressed in terms of $z_t$ as:
\begin{equation}
E\left[\|\boldsymbol z_{t+1}\| \big|\boldsymbol z_t,y_t\right]\leq \delta \|\boldsymbol z_t\|.
\label{eqUl2}
\end{equation}
Equation \eqref{gamisemas} shows that the both the right and the left hand side of  \eqref{eqUl2} are equal to the corresponding terms of\eqref{eqUl1}.
 Hence, it remains to prove that (ii) is equivalent to \eqref{eqUl2}.
 
 It holds:
 \begin{align*}
 E\left[\|\boldsymbol z_{t+1}\|\big |\boldsymbol z_t,y_t=i\right] &=F(\boldsymbol z_t,i)=\\&= \sum_{j=1}^{\bar{M}}  \sum_{i'=1}^M \tilde{c}(i,j,i') \vect 1^T \boxtimes \tilde{\boldsymbol A}(i',i,j) \boxtimes \boldsymbol z_t.
 \end{align*}
 The function $F(\boldsymbol z,i)$ is 1-homogeneous in $\boldsymbol z$. Thus, \eqref{eqUl2} is equivalent to 
 \begin{align*}
 \max_{\|\boldsymbol z\|\leq 1} F(\boldsymbol z,i) \leq \delta, \text{ ~  for } i=1,\dots,M.
 \end{align*}
 Furthermore, $F(\boldsymbol z,i)$ is non-decreasing in $\boldsymbol z$. Thus, (i) is equivalent to $ F(\vect 1,i) \leq \delta$, which is equivalent to (ii). \hfill $\square$
 
 \begin{remark}
 	Equation \eqref{Modif_cond} is stated using the matrix $\tilde A$, which is computed in transformed coordinates (equation \eqref{Alpha_tilda_eq}). A similar  transformation is used  (in a different context) in \cite{van2012modeling}, in order to define the `maximum autonomous growth rate'.
 \end{remark}

For the needs of the rest of the current section we shall use $k$ in the place of $t$,$A(y)$ in the place of  $\bar A(y,w)$ and $x$ in the place of $\bar x$.

 \begin{corollary}
 	\label{Corrol_Simpl_cond}
 	Assume that:
 	\begin{align*}
 	\sum_{j=1}^M c_{ij} \boldsymbol p(j)^T\boxtimes \boldsymbol A(i) \boxtimes (\boldsymbol p^{-1}(i)) \leq \delta,
 	\end{align*}
 	for $i=1,\dots,M$, $\delta<1$ and $\boldsymbol p^{-1}(i)$ is a vector having as entries the inverses of the entries of  $\boldsymbol p(i)$. Then the system  given by \eqref{Max_Prod_MJ} is mean norm exponentially stable and almost surely stable.
 \end{corollary}
 \textit{Proof:} Using $k$ in the place of $t$, $A(y)$ in the place of  $\bar A(y,w)$ and $x$ in the place of $\bar x$ in Lemma \ref{Lemma_1point} we get that the conditions of   Proposition \ref{Prop_lyap_Cond} hold true.  Thus, the system  given by \eqref{Max_Prod_MJ} is mean norm exponentially stable and almost surely stable.   \hfill $\square$

 We then consider Max-Product stochastic systems with inputs and outputs and the notion of BIBO stability in probability is introduced. 
 
 \begin{definition}
 	Consider the system:
 	\begin{align}
 	\boldsymbol x_{k+1}&=\boldsymbol A(y_k)\boxtimes \boldsymbol x_k \vee \boldsymbol B(y_k)\boxtimes \boldsymbol u_k  \label{BIBipO1},\\
 	\boldsymbol z_k &=\boldsymbol  C(y_k)\boxtimes \boldsymbol x_k  \label{BIBipO2}. 
 	\end{align}
 	The system is Bounded Input Bounded in probability Output (BIBipO) stable if for any $\varepsilon>0$, $M_u>0$ and any initial condition, there exist a bound $M_z>0$ such that:
 	\begin{align}
 	P(\|\boldsymbol z_k\|\leq M_z)>1-\varepsilon. \label{BIBipO_ineq}
 	\end{align}
 \end{definition}
 
 The following proposition shows that the exponential mean norm stability of the free system implies the BIBipO stability. 
 
 \begin{proposition}
 	\label{BIBipO_prop}
 	If the free system given by \eqref{Max_Prod_MJ} is mean norm exponentially stable then the system \eqref{BIBipO1}-\eqref{BIBipO2} is BIBipO stable.
 \end{proposition}
 \textit{Proof:} Consider a pair of constants $\varepsilon>0$, $M_u>0$ and an initial condition $\boldsymbol x_0\in \mathbb{R}^n_+$. Following \cite{Maragos_submited} the state vector can be written as:
 \begin{equation}
 \boldsymbol x_k = \boldsymbol\Phi(k,0) \boxtimes \boldsymbol x_0 \vee  \left (   \bigvee_{t=1}^k\boldsymbol\Phi(k,t)\boxtimes \boldsymbol B(y_{t-1})     \boxtimes \boldsymbol u_{t-1}  \nonumber                                \right ),
 \end{equation}
 where $\boldsymbol\Phi$ is the transition matrix given by:
 \begin{equation}
 \boldsymbol\Phi(k_2,k_1) = \begin{cases} \boldsymbol A(y_{k_2-1}) \boxtimes \dots \boxtimes \boldsymbol A(y_{k_1}) &\mbox{if } k_2>k_1 \\ 
 I & \mbox{if } k_2=k_1. \end{cases} \nonumber
 \end{equation}
 
 For any given constant $M_x>0$ it holds:
 \begin{align}
 &P[\|\boldsymbol x_k\|>M_x] \leq P [\|\boldsymbol\Phi(k,0) \boxtimes \boldsymbol x_0 \|>M_x]+\nonumber\\&~~~~+\sum_{t=1}^k P[\|\boldsymbol\Phi(k,t)\boxtimes \boldsymbol B(y_{t-1})     \boxtimes \boldsymbol u_{t-1} \|>M_X]\nonumber\\
 &~~~~\leq P[\|\boldsymbol\Phi(k,0) \boxtimes \boldsymbol x_0 \|>M_x]+\nonumber\\&~~~~+\sum_{t=1}^k P[\|\boldsymbol\Phi(k,t)\boxtimes \bar{ \boldsymbol U }    \|>M_x], \label{inToclaim}
 \end{align} 
 where $\bar {\boldsymbol U}=  \max \left\{  \| \boldsymbol B(i)\boxtimes \boldsymbol u\|: \|\boldsymbol u\|\leq M_u, i=1,\dots,M \right\}\vect{1}$.
 
 The following claim will be used:
 \newline
 \textit{Claim}: There exists a value $M_x>0$ such that the right hand side of the last inequality in \eqref{inToclaim} is less than $\varepsilon$ for any positive integer $k$.
 
 To prove the claim we first use the Markov inequality:
  \begin{align}
\label{fromMarkovIneq}
 &E[\|\boldsymbol x_k\|>M_x] \leq \frac{1}{M_x} \left [E[\|\boldsymbol\Phi(k,0) \boxtimes \boldsymbol x_0 \|]+\sum_{t=1}^k E[\|\boldsymbol\Phi(k,t)\boxtimes \bar{ \boldsymbol U }    \|] \right]. 
 \end{align} 
 The term 
 $E[\|\boldsymbol\Phi(k,0) \boxtimes \boldsymbol x_0 \|]$ is bounded, due to the  mean norm exponential stability of the free system. Then, observe that it holds $E[\|\boldsymbol\Phi(k,t)\boxtimes \bar{ \boldsymbol U }\|] =E[\|\tilde x_{k-t}\|]$ where  $\tilde x_l$ satisfies: 
 \begin{align}
 \tilde x_{l+1} &= A(y_{k-t+l})\tilde x_{l} \label{helping1},\\
 \nonumber \tilde x_0&= \vect{\bar U}.
 \end{align}
 The system \eqref{helping1} is mean norm exponentially stable. Thus:
 \begin{equation}
 \sum_{t=1}^k E[\|\boldsymbol\Phi(k,t)\boxtimes \bar{ \boldsymbol U }    \|] \leq \sum_{t=1}^\infty E[\|\boldsymbol\Phi(k,t)\boxtimes \bar{ \boldsymbol U }    \|] \leq \frac{aL}{a-1}\| \bar{ \boldsymbol U }    \|,
 \end{equation}
 where $a$ and $L$ the constants satisfying the mean norm exponential stability definition. Hence, the right hand side of \eqref{fromMarkovIneq} tends to zero as $M_x$ increases, which completes the proof of the claim.
 
  Hence, a constant $M_z$ satisfying \eqref{BIBipO_ineq} is given by: $M_z=\max \left\{ \|\boldsymbol C(i)\boxtimes \boldsymbol x\|: \|\boldsymbol x\|\leq M_x, i=1,\dots,M \right\}$. \hfill $\square$

 \section{$k$-Step Lyapunov Functions }
 \label{Many_Step_Lyapunov_Functions}
 In the last section, Lyapunov functions were used for the stability analysis of Max-Product systems with Markovian jumps. In this section we consider $k$-step Lyapunov functions and derive necessary and sufficient conditions for the mean-norm exponential stability. It turns out that many step Lyapunov functions offer greater flexibility. 
 
 We shall consider Lyapunov functions $V: \mathbb{R}_+^n\times \{1,\dots, M\} \rightarrow \mathbb{R}_+$  with the following properties:
 \begin{itemize}
 	\item[\textit{P1.}] $V(\boldsymbol x,y)$ is $1$-homogeneous in $\boldsymbol x$.
 	\item[\textit{P2.}] $V(\boldsymbol x,y)$ is continuous in $\boldsymbol x$.
 	\item[\textit{P3.}] It holds $V(\boldsymbol x,y)=0$ iff $\boldsymbol x=0$.
 \end{itemize}
 
 The following proposition gives necessary and sufficient conditions for the mean-norm exponential stability in terms of many step Lyapunov functions.
 
 \begin{proposition}
 	\label{Many_step_Lyap_Prop}
 	Consider a function $V(\boldsymbol x,y)$ satisfying (P1)-(P3). Then, the following are equivalent:
 	\begin{itemize}
 		\item[(i)] The system given by \eqref{Max_Prod_MJ} is mean-norm exponentially stable.
 		\item[(ii)] For each $\delta\in(0,1)$, there exists a positive integer $k_0$ such that:
 		\begin{equation}
 		\label{ManyStLyap_C1}
 		E[V(\boldsymbol x_k,y_k)]\leq \delta V(\boldsymbol x_0,y_0),
 		\end{equation}
 		for any $\boldsymbol x_0\in\mathbb{R}_+^n$, $y_0\in\{1,\dots, M\}$ and any $k\geq k_0$.
 		
 		\item[(iii)] There exists a $\delta\in(0,1)$ and a positive integer $k_0$ such that:
 		\begin{equation}
 		\label{ManyStLyap_C2}
 		E[V(\boldsymbol x_{k_0},y_{k_0})]\leq \delta V(\boldsymbol x_0,y_0),
 		\end{equation}
 		for any $\boldsymbol x_0\in\mathbb{R}_+^n$, $y_0\in\{1,\dots, M\}$.
 	\end{itemize}
 \end{proposition}
 \textit{Proof}: (i) $\Rightarrow$ (ii). The following claim is first proved:
 
 \textit{Claim}: There exist positive constants $b_{min}$ and $b_{max}$ such that: 
 \begin{equation}
 b_{min}\|\boldsymbol x\|\leq V(\boldsymbol x,y)\leq b_{max}\|\boldsymbol x\|\label{Lyap_FUN_bounds}.
 \end{equation}
 From (P2) and (P3) the values of the constants $b_{min}$ and $b_{max}$ defined by:
 \begin{align*}
 b_{min}=\min\{V(\boldsymbol x,y): \boldsymbol x\in \mathbb{R}^n_+, \|\boldsymbol x\|=1\},\\
 b_{max}=\max\{V(\boldsymbol x,y): \boldsymbol x\in \mathbb{R}^n_+, \|\boldsymbol x\|=1\},
 \end{align*}
 are finite and positive. Then, (P1) completes the proof of the claim. 
 
 Assume that the system given by \eqref{Max_Prod_MJ} is mean-norm exponentially stable and $a$ and $L$ satisfy Definition \ref{Part3Def1} part (iii).  Fix a $\delta\in(0,1)$. It holds:
 \begin{align*}
 E[V(\boldsymbol x_k,y_k)]]&\leq E[b_{max}\|\boldsymbol x_k\|] \leq\\&\leq b_{max} L \|\boldsymbol x_0\|/a^k\leq\frac{b_{max}}{b_{min}} L a^{-k} V(\boldsymbol x_0,y_0).
 \end{align*}
 Choosing $k_0$ such that $\frac{b_{max}}{b_{min}} L a^{-k_0}<\delta$, inequality \eqref{ManyStLyap_C1} is satisfied.
 
 (ii) $\Rightarrow$ (iii) is trivial.
 
 (iii) $\Rightarrow$ (i). Using the same arguments as in the first part of the proof it   is easy to see that there exists a positive integer $N_0$ such that:
 \begin{align*}
 E[\|\boldsymbol x_{N_0k_0}\|]\leq \delta \|\boldsymbol x_0\|.
 \end{align*}
 Consider the Euclidean division of $k$ by $N_0k_0$, i.e.  $k=(N_0k_0) q+r$. Using repeatedly the following inequality:
 \begin{align*}
 E[\|\boldsymbol x_k\|] =& E\left[E\left[\|\boldsymbol x_k\|\big|\boldsymbol x_{k-N_0k_0},y_{k-N_0k_0}\right]\right]\\&\leq
 \delta E[\|\boldsymbol x_{k-N_0k_0}\|],
 \end{align*} 
 we obtain:
 \begin{align}
 E[\|\boldsymbol x_k\|]\leq \delta^q E[\|\boldsymbol x_r\|]. \label{x_k_ineq}
 \end{align} 
 Furthermore, $r$ as a remainder satisfies $0\leq r< N_0k_0$ and $q$ as a quotient satisfies $q\geq \frac{k}{N_0k_0}-1$. 
 A bound for $E[\|\boldsymbol x_r\|]$ is then derived using repeatedly the following inequality:
 \begin{align}
 \|\boldsymbol A(y)\boxtimes \boldsymbol x\| \leq \left[\max_{i,j,y}A_{ij}(y)\right] \|\boldsymbol x\| \nonumber.
 \end{align}
 Inequality \ref{x_k_ineq} implies that:
 \begin{align}
 E[\|\boldsymbol x_k\|]\leq \delta^{\frac{k}{N_0k_0}} \left[\max_{i,j,y} A_{ij}(y)\right] ^{N_0k_0-1}/\delta. \nonumber
 \end{align} 
 Thus, using for $a$ and $L$ the values $a=\delta^{1/(N_0k_0)}$ and $L=\left[\max_{i,j,y} A_{ij}(y)\right] ^{N_0k_0-1}/\delta$, the inequality in Definition \ref{Part3Def1} part (iii) holds true and the system is mean norm exponentially stable. \hfill $\square$

 The following corollary uses Lyapunov functions in the form $V(\boldsymbol x,y) = p(\boldsymbol y)^T\boxtimes \boldsymbol x$ and Lemma \ref{Lemma_1point}. Particularly, a system in the form \eqref{x_bar_equation} is considered with $\bar {\boldsymbol x}_t=\boldsymbol x_{k_0t}$. 
 
 \begin{corollary}
 	Fix a positive integer $k_0$. Assume that there exists a set of vectors $\boldsymbol p(1),\dots,\boldsymbol p(M)$ such that:
 	\begin{align}
 	\sum_{(j_1,\dots,j_{k_0-1}),i'} &\tilde{c}(i,(j_1,\dots,j_{k_0-1}),i') \vect 1^T \boxtimes  \nonumber \\ &\boxtimes\tilde{\boldsymbol A}(i',i,(j_1,\dots,j_{k_0-1}),i')) \boxtimes \vect 1 \leq \delta, \label{simplifiedCorollCond}
 	\end{align}
 	where $\delta<1$, the matrix $\tilde {\boldsymbol A}$ is given by \eqref{Alpha_tilda_eq}, the matrix $\bar{\boldsymbol A}$ 
 	by:
 	\begin{align*}
 	\bar{\boldsymbol A}(y_t,(j_1,\dots,j_{k_0-1}),i')=\boldsymbol A(j_{k_0-1})\boxtimes \dots \boxtimes \boldsymbol A(j_1),
 	\end{align*}
 	and the constants $\tilde{c}$ by:
 	\begin{align*}
 	\tilde{c}(i,(j_1,\dots,j_{k_0-1}),i') = c_{ij_1}c_{j_1j_2}\cdot\dots\cdot c_{j_{k_0-1}i'}.
 	\end{align*}
 	Then \eqref{Max_Prod_MJ} is mean norm exponentially stable. Furthermore, if \eqref{Max_Prod_MJ} is mean norm exponentially stable then there exists a positive integer $k_0$ and a set of vectors $\boldsymbol p(1),\dots,\boldsymbol p(M)$ satisfying \eqref{simplifiedCorollCond}. \hfill $\square$
 \end{corollary}

 \section{Max-Plus Systems with Markovian Jumps}
 \label{Max_plus_MJ_Section}
 
 \subsection{Almost Sure Bounds for the Free System}
 We then turn to Max-Plus systems with Markovian Jumps in the form 
 \eqref{Max_Plus_MJ}. An almost sure bound  on the evolution of the state of \eqref{Max_Plus_MJ} will be derived using the results of the previous sections. 
 
 For a given system in the form \eqref{Max_Plus_MJ} and a positive constant $\gamma$, we construct an equivalent Max-Product system. Particularly, consider the vector ${\boldsymbol x}'_k = \exp(\boldsymbol x_k)/\gamma^k$, where the exponentiation is considered component-wise. Then, ${\boldsymbol x}'_k$ evolves according to:
 \begin{align}
 {\boldsymbol x}'_{k+1}&={\boldsymbol A}'(y_k)\boxtimes {\boldsymbol x}'_k\label{Max_Prod_MJ2},\\
 {\boldsymbol x}'_0&\in\mathbb{R}^n_+ \nonumber,
 \end{align}
 and ${\boldsymbol A}'=\exp(\boldsymbol A)/\gamma$ where  the exponentiation is again considered component-wise.
 
 \begin{remark}
 	A transformation of a Max-Plus system to a sub-linear system is used in \cite{chang1996exponentiality}.  In contrast to the transformation   to a sub-linear system the transformation to a max-product system is exact (invertible). Let us note that the proof of the following proposition uses similar techniques with the proof of Corollary 2.3 of \cite{chang1996exponentiality}.
 \end{remark}

 The mean norm exponential stability of \eqref{Max_Prod_MJ2} can be used to derive some almost sure bounds for the evolution of \eqref{Max_Plus_MJ}.
 
 \begin{proposition}
 	\label{Max_plus_stab}
 	Assume that \eqref{Max_Prod_MJ2} is Mean-norm exponentially stable. Then almost all the sample paths of \eqref{Max_Plus_MJ} satisfy:
 	\begin{equation}
 	\boldsymbol x_k <( k \ln \gamma )\vect 1,
 	\end{equation}
 	for large $k$.
 \end{proposition}
 
 \textit{Proof:}  Consider the sets:
 \begin{equation}
 B_k =\{\omega\in \Omega : \boldsymbol x_k \nless( k \ln \gamma )\vect 1 \}.
 \end{equation}
 
 It holds $E[\|{\boldsymbol x}'_k\|] \leq M /a^k$ for some $a>1$. 
 Thus, using Markov inequality $P[\|{\boldsymbol x}'_k\|>1]\leq M /a^k$. Furthermore, $\|{\boldsymbol x}'_k\|>1$ iff $\boldsymbol x_k \nless ( k \ln \gamma )\vect 1$. Hence, $P(B_k)\leq M/a^k$ and $\sum_{k=1}^\infty P(B_k)<\infty$. 
 
 Thus, 1st Borel-Cantelli lemma (eg. \cite{billingsley2008probability}) applies. Hence:
 \begin{equation}
 P(\lim\sup B_k) =0,
 \end{equation}
 which concludes the proof. \hfill $\square$
 
 The results of Proposition \ref{Max_plus_stab} can be used to bound the Lyapunov Exponent of a Max-Plus systems with Markovian jumps. Conditions for the existence of the Lyapunov exponent are given in 
 \cite{cohen1988subadditivity}.
 
 \begin{corollary}
 	Assume that the system described by \eqref{Max_Plus_MJ} has a Lyapunov exponent $\ell$. Furthermore, assume that   \eqref{Max_Prod_MJ2} is mean norm exponentially stable. Then, $\ell<\ln \gamma$.
 \end{corollary}
 \textit{Proof}: It holds $\boldsymbol x_k/k<\gamma \boldsymbol 1$ for large $k$, almost surely. \hfill $\square$

 \subsection{Systems with Inputs}
 
 In this section systems of the form: 
 \begin{align}
 \boldsymbol x_{k+1}&=\left(\boldsymbol A(y_k)\boxplus \boldsymbol x_k \right )\vee \left(\boldsymbol B(y_k)\boxplus  u_k\right )  \nonumber, \\
 \boldsymbol z_k &=\boldsymbol  C(y_k)\boxplus \boldsymbol x_k,   
 \end{align}
 are considered in the context of the multi machine production system example studied in the following section. Using the results of Theorem 1 of  \cite{van2012modeling}, we assume that the input signal $u_k$ is scalar and that it grows in an approximately linear fashion:
 \begin{equation}
 u_k=kT+\delta_k, \label{giamiesaimalakarev}
 \end{equation}
 with $\delta_k$ bounded and  $T$ a positive constant. In \cite{van2012modeling} it is proved that, under certain additional conditions,   inputs in the form \eqref{giamiesaimalakarev} stabilize the corresponding switching Max-Plus linear system. 
 
 The following proposition shows that the difference of the state vector entries from $kT$ are bounded in probability. Let us note that the boundedness of these differences have been used in the literature to define a notion of stability for discrete-event systems \cite{van2012modeling}, \cite{passino1998stability}.
 
 \begin{proposition}
 	\label{Throughput_prop}
 	If the system \eqref{Max_Prod_MJ2} with $\gamma=e^T$, where $e$ is the basis of the natural logarithm, is mean norm exponentially stable then for any $\varepsilon>0$ there exists a bound $M_x$ such that:
 	\begin{equation}
\label{stoch_Stab_MAXplus}
 	P[x_k^i-kT\leq M_x]>1-\varepsilon,
 	\end{equation}
 	for any $k$, where $x_k^i$ is the $i$-th component of the vector $\boldsymbol x_k$.
 \end{proposition}
 \textit{Proof}: Consider the vector  ${\boldsymbol x}'_k = \exp(\boldsymbol x_k)/\gamma^k=\exp(\boldsymbol x_k)/\exp{kT}$. This vector evolves according to:
 \begin{align}
 {\boldsymbol x}'_{k+1} =\left(  {\boldsymbol A}'(y_k)\boxtimes {\boldsymbol x}'_k\right )\vee\left({\boldsymbol B}'(y_k)\boxtimes d_k \label{pr7eq1}\right ),
 \end{align}
 where $d_k=e^{\delta_k}$, ${\boldsymbol A}'=\exp(\boldsymbol A)/\gamma$ and ${\boldsymbol B}'=\exp(\boldsymbol B)/\gamma$ where all the matrix exponentiations are considered component-wise. Then, the application of Proposition \ref{BIBipO_prop}  to \eqref{pr7eq1} competes the proof. \hfill $\square$
 
 \section{Numerical Examples}

\subsection{Deterministic Max-Product Systems }

In this section we use the Lyapunov analysis of deterministic max-product systems to analyze slightly `nonlinear' max-plus systems. Such systems may arise in the modeling of discrete event systems for which the transport, processing, holding or idle times depend on  system operation. For example, the necessary cooling time for a machine in a production system may depend on the length of the previous cycle. Another example is the loading or boarding times in a rail transportation system which depend on the quantity of products or the number of passengers waiting to be served, which in turn may depend on the length of the last cycle. In this section we analyze a simple model of such a discrete event system.

Consider the two dimensional model:
\begin{align} x^1_{k+1}= \max\left(x^1_k+\bar a_{11},x^2_k +\bar a_{12}\right)+f_1(x^1_{k+1}-x^1_{k})\nonumber,\\x^2_{k+1}= \max\left(x^1_k+\bar a_{21},x^2_k+\bar a_{22}\right)+f_2(x^2_{k+1}-x^2_{k}),\end{align}
where $x^i_k$ represents the instant of time at which an event takes place for the $k-$th time (eg. the train departs from station $i$) and   $f_1$, $f_2$ terms represent the  dependence on the length of the last cycle. For simplicity assume that $f_1$ and $f_2$ are linear: $f_1(z)=z_2(z)=\bar\delta z$, with $|\bar \delta|<1/2$.

Then (1) can be be written as:
\begin{align} x^1_{k+1}= \max(x^1_k+a_{11},x^2_k +a_{12}-\delta(x^1_k-x^2_k)),\nonumber\\
x^2_{k+1}= \max(x^1_k+a_{21}-\delta(x^2_k-x_k^1) ,x^2_k+a_{22}),
\label{mpnonlin1}
\end{align}
where $a_{ij}=a_{ij}/(1-\delta)$ and $\delta=\bar\delta/(1-\delta)$. This system is clearly \textbf{not} of the max-plus form. 
In order to analyze \eqref{mpnonlin1}, consider the corresponding exponentiated system: 
\begin{align} 
x'^1_{k+1}=& \max\left(a'_{11} x'^1_k,a'_{12}x'^2_k (x'^1_k/x'^2_k)^{-\delta}\right),\nonumber\\
x'^2_{k+1}= &\max\left(a'_{21}x'^1_k(x'^2_k/x'^1_k)^{-\delta} ,a'_{22}x'^2_k \right)
\label{mprnonlin1},
\end{align}
where $x'^i_k = \exp(x^i_k)/\gamma^k $, $a'_{ij}=\exp(a_{ij})/\gamma$. The dynamics \eqref{mprnonlin1} can be written as:
\begin{align} 
\boldsymbol x'_{k+1}=\boldsymbol A'(x'^1_k/x'^2_k)\boxtimes \boldsymbol x'_k,
\end{align}
where 
$$\boldsymbol A'(x'^1_k/x'^2_k)=\left[\begin{matrix}
   a_{11}       & a'_{12} (x'^1_k/x'^2_k)^{-\delta}  \\
    a'_{21} (x'^2_k/x'^1_k)^{-\delta}       & a'_{22}
\end{matrix}\right].$$

Then, the stability of the dynamics \eqref{mprnonlin1} can be studied using the Lyapunov function of the max-product `linearized' system:
\begin{align} 
\boldsymbol x'_{k+1}=\boldsymbol A'(1)\boxtimes \boldsymbol x'_k.
\end{align}

\begin{example}
Assume that $a_{11}=a_{22}= 1.5686$, $a_{12}=1.7918$, $a_{21}= 1.3350
$, $d=-0.15$ and $\gamma=5$. 
 Then, the exponentiated system is:
\begin{align} 
x'^1_{k+1}=& \max\left(0.96 x'^1_k, 1.2 (x'^1_k/x'^2_k)^{0.15} x'^2_k\right ),\nonumber\\
x'^2_{k+1}= &\max\left(0.76x'^1_k(x'^2_k/x'^1_k)^{0.15} ,0.96x'^2_k \right),
\label{eq47_7}
\end{align}
and the  `max-product linearized' matrix is: 
$$\boldsymbol A'(1)=\left[\begin{matrix}
0.96 &   1.2\\
    0.76 &   0.96
\end{matrix}\right].$$
Using \eqref{LYap_form_min1}, \eqref{LYap_form} we obtain a  Lyapunov function for the  max-plus linearized system:
$$ V(\boldsymbol x) = [1~~1.25]\boxtimes \boldsymbol x.$$

 We then use $V$ as a Lyapunov function candidate for \eqref{eq47_7}. The function $f(\boldsymbol x) = A'(x^1/x^2)\boxtimes \boldsymbol x$ is 1-homogenus. Therefore, we need only to show that if $V(\boldsymbol x'_k)\leq 1$ implies $V(\boldsymbol x'_{k+1})\leq 1$. 
Equivalently we need to show that $x'^1_{k+1}\leq 1$ and $x'^2_{k+1}\leq 0.8$, if $x'^1_{k }\leq 1$ and $x'^2_{k }\leq 0.8$. Indeed for such  $\boldsymbol x'_k$ it holds: 
\begin{align}
x'^1_{k +1}&=\max\left(0.96 x'^1_k , 1.2 (x'^1_k)^{0.15} (x'^2_k)^{0.85}\right)\leq 1,\nonumber\\\nonumber
x'^2_{k +1}&=\max\left(0.76 (x'^1_k)^{0.85} (x'^2_k)^{0.15} , 0.96 x_k^2\right)\leq 0.8.
\end{align}
Thus,  the system \eqref{eq47_7} is stable. \hfill $\square$
\end{example}
Let us note it is not possible to analyze \eqref{mpnonlin1} using directly max-plus techniques and the transformation to a max-product system is essential.

 \subsection{Max-Product Systems with Markovian Jumps}

 In this section, we present a very simple numerical example
 of a Max-Product system with Markovian jumps. The Markov chain has two possible states $y_k\in\{1,2\}$ and the values of  matrix $\boldsymbol {A}$ are given by:
 \begin{align}
 \label{SystemOfExample}
 {\boldsymbol A}(1)=\left[\begin{matrix}
 1.05  &   1.5      \\
 0.4  &  0.3
 \end{matrix} \right], {\boldsymbol A}(2)=\left[\begin{matrix}
 0.5 &   0.4      \\
 0.7 &  0.3
 \end{matrix} \right],
 \end{align}
 and the Markov chain has transition probability matrix:
 \begin{align*}
 {\boldsymbol c}=\left[\begin{matrix}
 0.3  &   0.7      \\
 0.4  &  0.6
 \end{matrix} \right].
 \end{align*}
 
\begin{figure}
 	\centering
 	\includegraphics[width=3.6in]{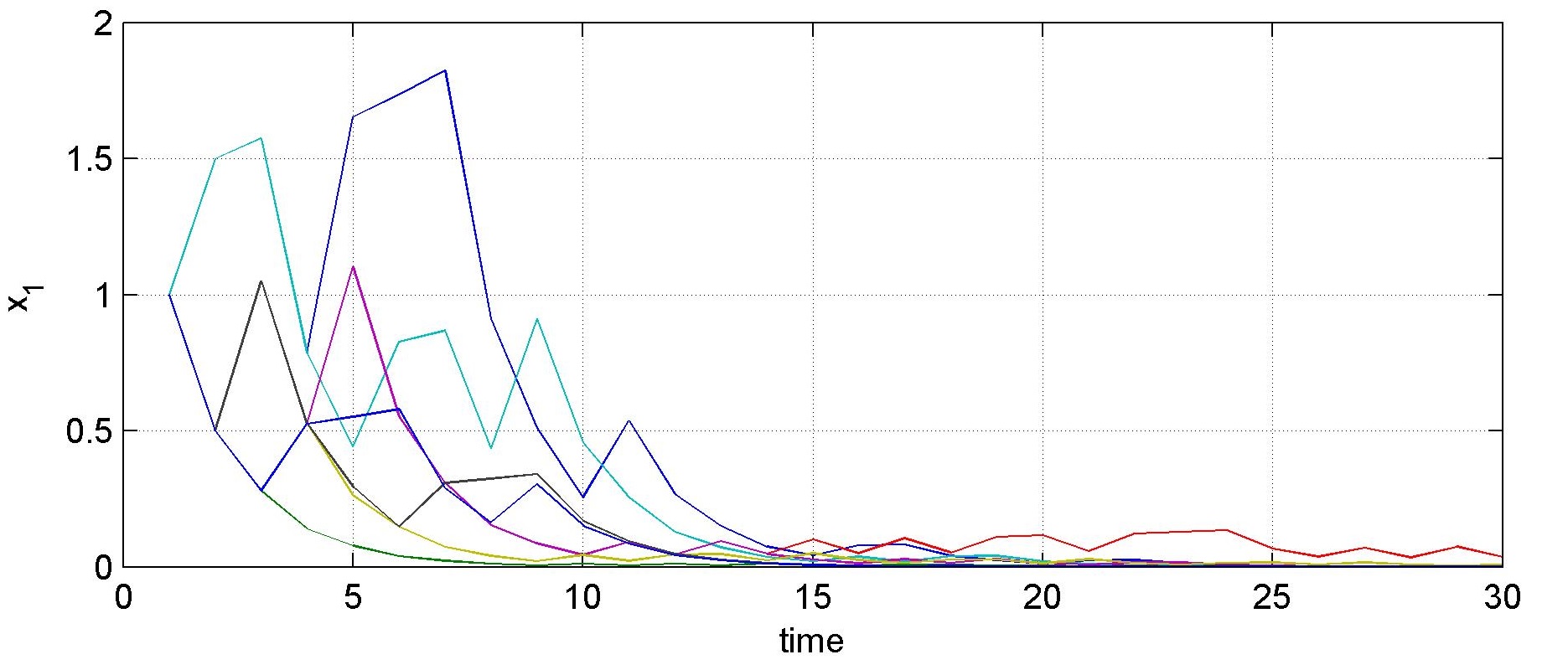}
 	\centering
 	\includegraphics[width=3.6in]{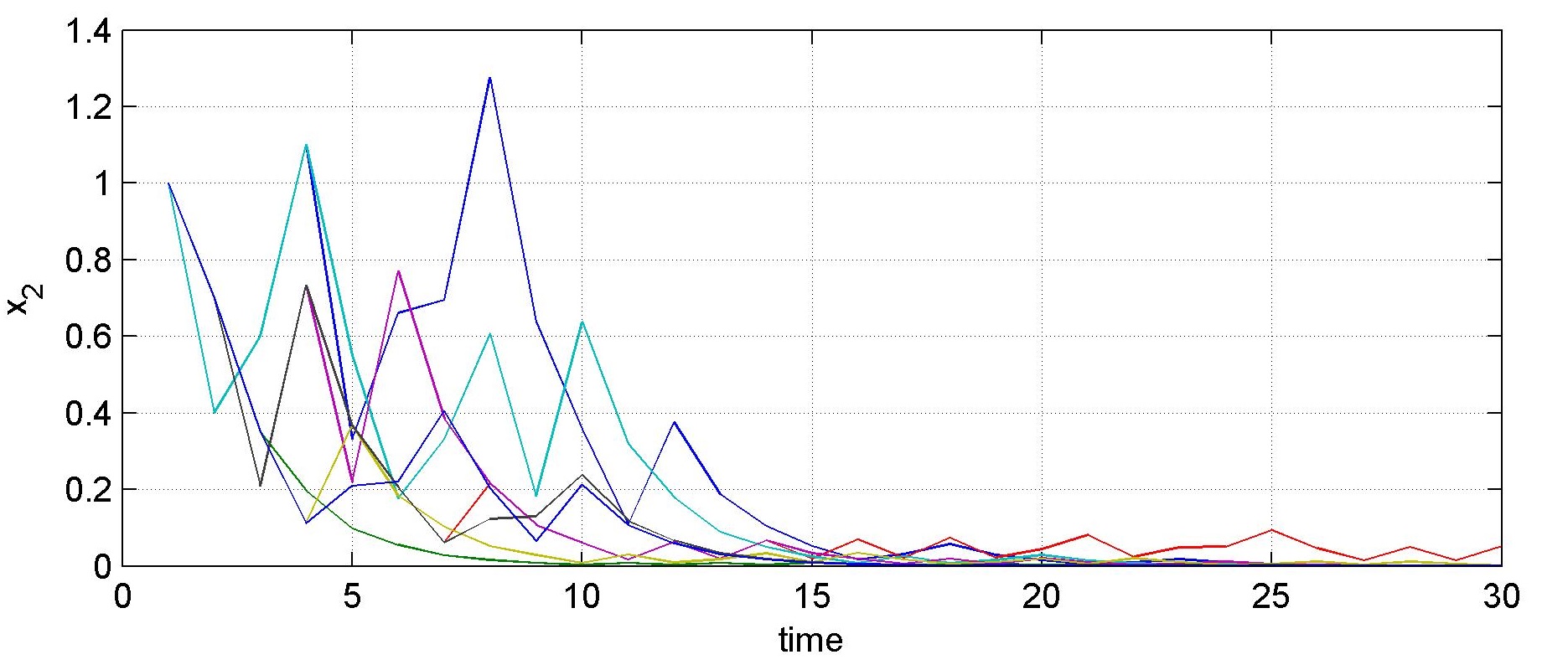}
 	\caption{Several sample paths of the system described by \eqref{SystemOfExample}. (Best viewed in color)}
 	\label{SamplePaths}
 \end{figure}

 Using simple search techniques a Lyapunov function satisfying the conditions of Corollary \ref{Corrol_Simpl_cond} can be
 obtained. One of those Lyapunov functions is:
 \begin{align*}
 \boldsymbol p(y) = \begin {cases} [4 ~6]^T & \mbox{if } y=1\\ 
 [3 ~2]^T & \mbox{if } y=2. \end{cases}   
 \end{align*}
 Hence, the system is mean norm exponentially stable. Several sample paths of the system are given in Figure \ref{SamplePaths}.          
 
 \subsection{Application  to Multi-Machine Production Systems}
 \label{NUM_EX_SEC}
\begin{figure}
 	\centering
 	\includegraphics[width=3.6in]{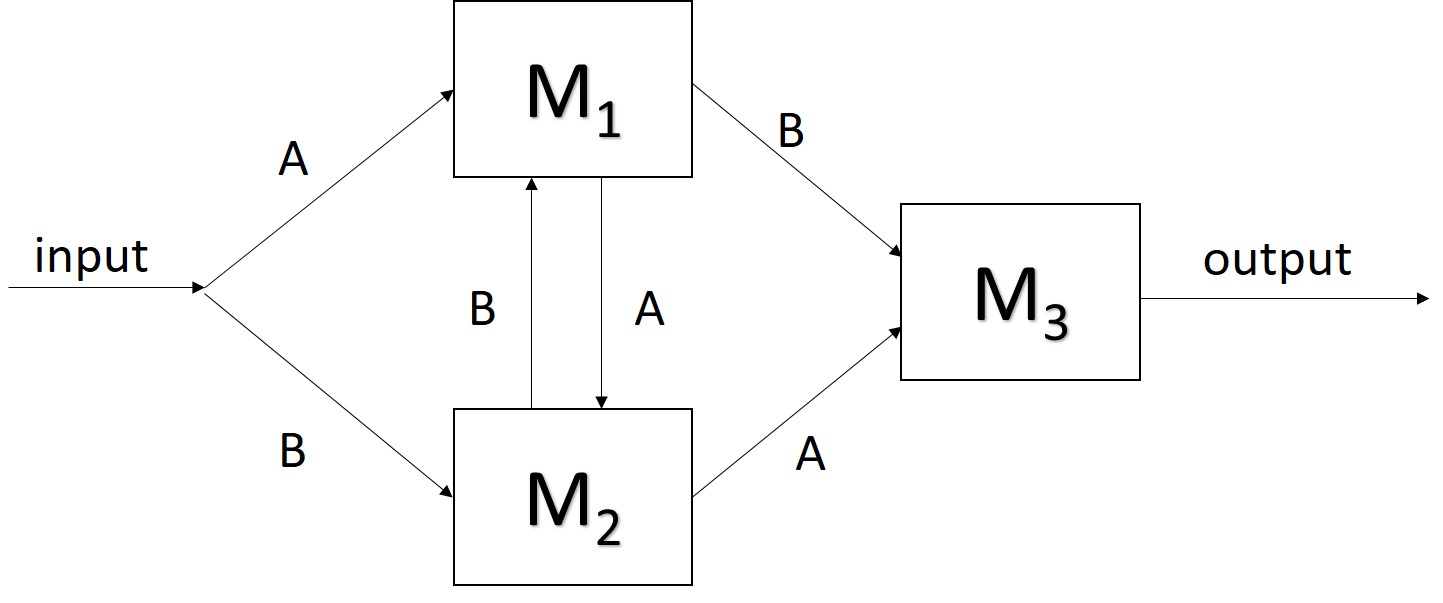}
 	\caption{The production system}
 	\label{ProductionSytem}
 \end{figure}

 In this section we study a very simple example of a production system consisting of three machines analyzed in \cite{van2012modeling}. The production system may produce two distinct outputs A and B. The order in which the machines process the raw material is different for the two products. Particularly, when the product A is produced, the machines are used with order $M_1\rightarrow M_2\rightarrow M_3$ while when the product B is produced the order is $M_2\rightarrow M_1\rightarrow M_3$. The production system is depicted in Figure \ref{ProductionSytem}. An important question is that of maximum throughput. Maximum throughput  is  the maximum rate at which the system can process the raw material and it is defined as the inverse of the minimum cycle time (\cite{baccelli1992synchronization}).
 
 A max-plus stochastic system describing the timing of the production system will be described. Each machine starts working as soon as possible, that is when the input material is available and also it has finished all the previous work. Let us denote by $u_k$ the time instant at which the raw material for the $k-th$ product becomes available and by $x_k^i$ the time instant at which the machine $i$ starts working for the production of the $k$-th product. We assume that the processing time for the machines are $s_1=1$, $s_2=2$ and $s_3=1$. Furthermore, $z_k$ denotes the time instant at which the product $k$ becomes available. 
 
 The evolution of $\vect{x}_k$ and $\boldsymbol z_k$ is given by:
 \begin{align}
 \boldsymbol x_{k+1}&=\left(\boldsymbol A(y_k)\boxplus \boldsymbol x_k\right) \vee \left(\boldsymbol B(y_k)\boxplus  u_{k+1} \right), \nonumber \\
 \boldsymbol z_k &=\boldsymbol  C\boxplus \boldsymbol x_k,  
 \label{Prod_SYS_eq} 
 \end{align}
 where $y_k=1$ when the product A is produced and $y_k=2$ when product B is produced. The matrices $\boldsymbol A(1),\boldsymbol A(2), \boldsymbol B(1),\boldsymbol B(2)$ and $\boldsymbol C$ are given by:
 \begin{equation}
 \boldsymbol A(1)=\left[\begin{matrix}
 s_1& -\infty & -\infty\\
 2s_1& s_2 & -\infty\\
 2s_1+s_2& 2s_2 & s_3
 \end{matrix} \right], ~~\nonumber
 \boldsymbol B(1)=\left[\begin{matrix}
 0\\
 s_2\\
 s_1+s_2
 \end{matrix} \right],
 \end{equation}
 \begin{equation}
 \boldsymbol A(2)=\left[\begin{matrix}
 s_1& 2s_2 & -\infty\\
 -\infty& s_2 & -\infty\\
 2s_1& s_1+2s_2 & s_3
 \end{matrix} \right], ~~\nonumber
 \boldsymbol B(2)=\left[\begin{matrix}
 s_2\\
 0\\
 s_1+s_2
 \end{matrix} \right],
 \end{equation}
 and $\boldsymbol C=[-\infty ~~-\infty ~~s_3]$. The details can be found in \cite{van2012modeling}. 
 
 We assume that which product is produced at each time step depends on exogenous orders which behave randomly. Particularly we assume that $y_k$ is a Markov chain with transition probability matrix:
 \begin{equation}
 \boldsymbol c=\left[\begin{matrix}
 0.8& 0.2\\
 0.2&0.8
 \end{matrix} \right].\nonumber
 \end{equation}
 
 We further assume that the raw material arrives at a constant rate. Thus, the input signal has the form $u_k=kT$. Proposition \ref{Throughput_prop} will be used to show that the raw material that has arrived to the production system but not yet fully processed remains bounded in probability. Thus the system is capable of processing the raw material at the given rate. The condition that all the buffer levels remain bounded has been used in the literature to define the stability of Discrete Event Systems (eg. \cite{van2012modeling}).

\begin{figure}
 	\centering
 	\includegraphics[width=3.6in]{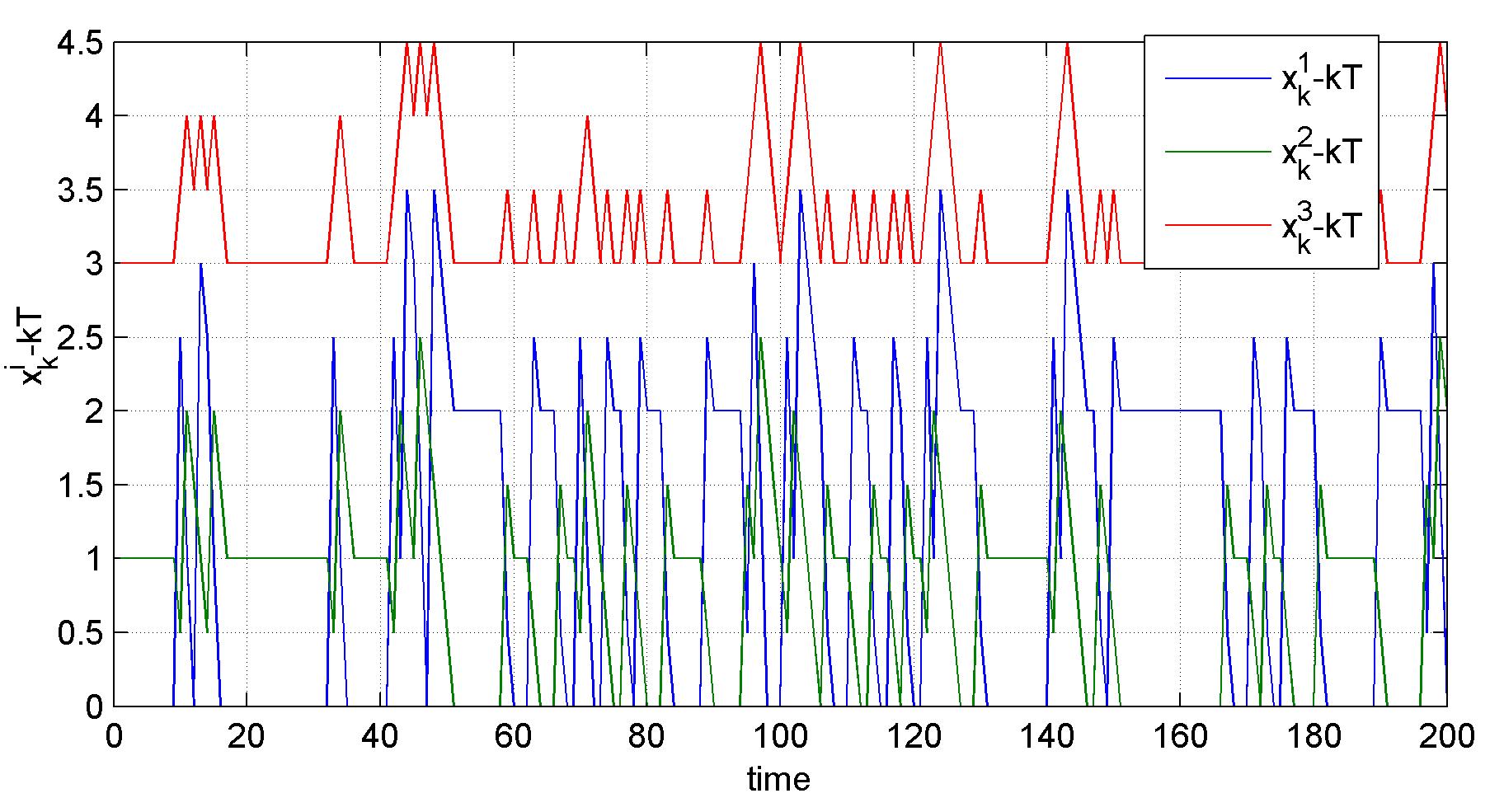}
 	\caption{The differences $x_k^i-kT$ in a sample path of the system \eqref{Prod_SYS_eq}. (Should be viewed in color).}
 	\label{Time_differece}
 \end{figure}

 \begin{example}
 	Assume that the raw material arrival has a period $T=2.5$. The matrices $ {\boldsymbol A}'(y_k)$ are: 
 	\begin{align*}
 	{\boldsymbol A}'(1)=\left[\begin{matrix}
 	0.2231  &   0      &   0\\
 	0.6065  &  0.6065   &      0\\
 	4.4817  &  4.4817   & 0.2231
 	\end{matrix} \right],\\
 	{\boldsymbol A}'(2)=\left[\begin{matrix}
 	0.2231 &   4.4817  &       0\\
 	0  &  0.6065   &      0\\
 	0.6065  & 12.1825   & 0.2231
 	\end{matrix} \right].
 	\end{align*}
 	The vectors $\vect{p}_1 =[12~ 12~ 1]$,  $\vect{p}_2 =[3~ 32~ 1]$ satisfy the conditions of Corollary  \ref{Corrol_Simpl_cond}. Hence, Proposition \ref{Throughput_prop} applies and 
 	$x_k^i-kT$ remains bounded in probability. 
 	Figure \ref{Time_differece} illustrates the evolution of stock times. 
 	\hfill \end{example}
 
 \begin{remark}
 	A stability condition is also derived in \cite{van2012modeling}. This stability condition  resembles the stability under arbitrary switching property. It turns out that, in contrast to usual linear systems, the stability under arbitrary switching property is easier to check than the stochastic stability in the Max-Plus systems (\cite{blondel2000approximating}). 
 	
 	The minimum value for $T$ satisfying the stability conditions of \cite{van2012modeling} can be computed using Linear Programming and for the current example has a value $T=3$. 
 	
 	Thus, the stochastic stability conditions \eqref{stoch_Stab_MAXplus} are less restrictive and allow the system to operate at a higher rate, compared with the stability under arbitrary switching. 
 \end{remark}
 
 \section{Conclusion}
 \label{CONCL_SECT}

 Max-Plus and Max-Product systems with Markovian jumps were  considered. A Lypaunov function is constructed for asymptotically stable deterministic Max-Product systems. This Lyapunov function is found to have a simple form and the stability conditions derived can be checked using Linear Programming. Slightly modified Lyapunov functions are then used to derive sufficient conditions for the mean norm exponential stability of Max-Product systems with Markovian Jumps. A simpler form of these conditions can be derived based on the monotonicity of the Lyapunov functions. Necessary and sufficient conditions for the mean norm exponential stability are then derived using many step Lyapunov functions. 
 
 Bounds for the evolution of the state of Max-Plus systems with Markovian jumps are  then derived, based on the results for the Max-Product systems. Finally a numerical example illustrates the application of the methods described on a production system.





~

\large{\textbf{References }}
\bibliographystyle{elsarticle-harv} 
\bibliography{refs}
\end{document}